# Universal behaviour of the SU(2) running coupling constant in the continuum limit *

M. Guagnelli[a]

[a] Dipartimento di Fisica, Università di Roma, Tor Vergata, and INFN, Sezione di Roma II,
Viale della Ricerca Scientifica, 00133 Roma, Italy

We present data from the ALPHA Collaboration about lattice calculation of SU(2) pure–gauge running coupling constant, obtained with two different definitions of the coupling itself, which show universality of the continuum limit and clarify the applicability of renormalized perturbation theory.

## 1. INTRODUCTION

Computation of $\alpha_{strong}$ by means of the finite-size technology of ref. [1] have been previously performed in the pure SU(2) and SU(3) gauge theory [2-7]. Using the massive computer power provided by the APE machines at Rome II and DESY, we were able to extend these calculations in several directions. Here and in a much more detailed manner in ref. [8], we discuss such developments.

## 2. THE TWO COUPLINGS

We will not enter into too many details, referring the interested reader to [8]. We will only recall that the Schrödinger functional involves an integration over all gauge fields with spatial periodic boundary conditions and fixed boundary conditions in time: it may be seen as the propagation kernel for going from the initial gauge field configuration $C$ at time 0 to the final configuration $C'$ at time $L$. Taking the derivative of the effective action respect to a parameter $\eta$ which enter in the definitions of classical boundary fields $C$ and $C'$, we can define a renormalized coupling which we call $\alpha_{SF}$. The only physical scale in the system is the box size $L$, so we can consider thw R.G. flow of the coupling arising from a change in this scale.

The other coupling is defined as the ratio of correlations of Polyakov loops, in a box with twisted-periodic boundary conditions. The correlations are taken in the time direction at a distance $L/2$, to ensure that the only scale entering in the definition is once again $L$. Correlations of Polyakov loop winding in one of the two twisted directions start in perturbation theory as $g_0^2$, while those in the periodic directions start as 1. Ratio of such correlations, with a normalization prefactor coming from a tree–level perturbative computation, allow us to define $\alpha_{TP}$.

At this point we would stress the fact that the infrared behaviours of $\alpha_{SF}$ and $\alpha_{TP}$ are completely different since the latter diverges exponentially in the $q \to 0$ limit, while the former is expected to approach a computed constant.

For both couplings we performed a perturbative computation giving us the relation between the coupling themselves and other more usual definition: for example

$$\alpha_{\overline{MS}} = \alpha_{SF} + 0.9433 \times (\alpha_{SF})^2 + \ldots \qquad (1)$$

with a second order coefficient which is under computation, and

$$\alpha_{\overline{MS}} = \alpha_{TP} - 0.5584 \times (\alpha_{TP})^2 + \ldots \qquad (2)$$

## 3. SIMULATIONS

In our simulations we used a standard hybrid over-relaxation algorithm with $N_{or}$ microcanonical sweeps followed by one heat–bath sweep. In all cases the high-quality random number generator of ref. [9] was used.

---

Surprisingly enough, we found that the integrated autocorrelation times of our observables depend strongly on the order in which each point of the lattice is visited during one sweep. The optimal choice was to update all links in a certain direction $\mu$ for all lattice points, then switching to the next value of $\mu$ and so on. The other choice, *i.e.* updating all four positive-directed links of a site before passing to the next site, results in a slightly more optimized code but in a much higher autocorrelation time (a factor 4 for SF and a factor 2 for TP, on moderately-sized lattices).

By studying autocorrelation times and relative variances of our observables, we find that the overall computational effort scales roughly as $(L/a)^6$ when one approaches the continuum limit at fixed $L$ for both couplings, with a bigger coefficient (about one order of magnitude) for TP case.

## 4. EXTRAPOLATION TO THE CONTINUUM LIMIT

The precision that can be reached for the two couplings at a given value of $L/a$ is quite different, and large lattices are very expensive for $\alpha_{TP}$. On the other hand lattice artifact are also different in the two cases: in particular for $\alpha_{TP}$ the leading corrections are expected to be $\mathcal{O}((a/L)^2)$, while for $\alpha_{SF}$ the presence of "source-walls" at the time edges of the lattice allows for terms $\mathcal{O}(a/L)$.

In TP case we found that not–too–big lattices (up to $16^4$) are enough for a safe extrapolation to the continuum limit, while in SF case we found that the use of Symanzik improvementent was of some help in reducing $\mathcal{O}(a/L)$ lattice artifact.

We consider also a perturbative improvement of the observables. For SF case we can build a 2-loops improved observable

$$\Sigma^{(2)}_{SF}(2,u,a/L) \equiv \frac{\Sigma_{SF}(2,u,a/L)}{1+\delta_1(a/L)u+\delta_2(a/L)u^2} \quad (3)$$

where $\Sigma^{(2)}_{SF}(2,u,a/L)$ is the lattice step scaling function with a scale factor $s = 2$ computed at $g^2_{SF} = u$, and $\delta_i(a/L)$ are perturbative corrections. For TP case we can only build the 1-loop improved observable.

This improvement is indeed very efficient in reducing the lattice artifacts for $\Sigma^{(2)}_{SF}$, while for $\Sigma^{(1)}_{TP}$ at the two largest value of the coupling the non-perturbative effects start to win over the improvement, leading to higher values of lattice artifacts.

## 5. RENORMALIZED PERTURBATION THEORY

Let us now try a quantitative comparison between the two renormalized couplings: we want to obtain a non-perturbative universal relation between $\alpha_{SF}(q)$ and $\alpha_{TP}(q)$. To achieve this goal, we decided to fix $\alpha_{SF}(L) = 0.16535$ for several values of $a/L$, and at the corresponding values of $\beta$ we compute $\alpha_{TP}(L)$. Then we can safely extrapolate to the continuum limit, keeping only the points for $L/a \geq 8$ and allowing for a linear dependence on $a/L$. In this way we obtain

$$\alpha_{TP}(q) = 0.2374(26) \quad at \quad \alpha_{SF}(q) = 0.16535 \quad (4)$$

At this point we can see how perturbation theory can account for this big difference: we know from eq. (1) and eq. (2) that the perturbative relation between $\alpha_{SF}$ and $\alpha_{TP}$ has a rather large 1-loop coefficient:

$$\alpha_{TP} = \alpha_{SF} + 1.5017 \times (\alpha_{SF})^2 + \ldots \quad (5)$$

It is clear that a naive application of this truncated expansion could give a not too satisfactory results for the range of coupling we have measured. In fact plugging in eq. (5) the value $\alpha_{SF} = 0.16535$ we get for $\alpha_{TP}$ the value $0.2064$, instead of the simulation result given in eq. (4).

However from previous analyses we know that both couplings follow the R.G. flow with the perturbative two–loops $\beta$–function in most of the range of scales condidered: this observation can easily bring to the conclusion that a perturbative relation between $\alpha_{TP}(q)$ and $\alpha_{SF}(fq)$ (note the shifted scale) may give better results provided the scale factor $f$ is choosen with care. An intuitive choice for $f$ is the ratio of $\Lambda$-parameters,

$$f = \Lambda_{SF}/\Lambda_{TP} = 0.27620(2) \quad (6)$$

because with this choice the perturbative relation reads

$$\alpha_{TP}(q) = \alpha_{SF}(fq) + \mathcal{O}((\alpha_{SF}(fq))^3) \quad (7)$$

If we now consider the scale $q_1$ such that $\alpha_{SF}(q_1) = 0.16535$, we get, interpolating our data using the effective 3-loops $\beta$-function,

$$\alpha_{SF}(fq_1) = 0.2289(6) \qquad (8)$$

which is indeed very close to the measured value of $\alpha_{TP}$ given in eq. (4). By applying this shift of the scale to all the measured points, we obtain figure 1, in which we can see the two couplings nicely superimposing each-other.

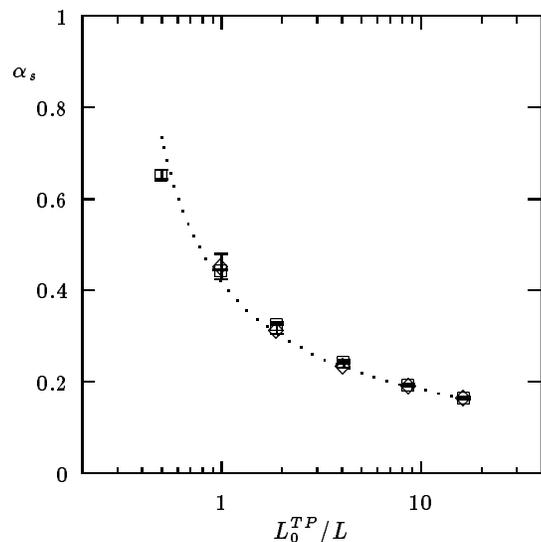

Figure 1. R.G. flow of the two couplings: scale of $\alpha_{SF}$ (diamonds) is shifted. Dotted curve is the two-loops universal evolution.